\documentclass[superscriptaddress,amsmath, nofootinbib]{revtex4}
\usepackage{amsfonts}
\usepackage{graphicx}
\usepackage[brazil]{babel}
\usepackage[latin1]{inputenc}
\usepackage{amsmath}
\usepackage{amssymb}
\begin{document}


\title{Comment on the Letter  "Geometric Origin of the Tennis Racket Effect''     
by P. Mardesic, et al, Phys. Rev. Lett. 125, 064301 (2020). }

\author{Alexei A. Deriglazov }
\email{alexei.deriglazov@ufjf.br} \affiliation{Depto. de Matem\'atica, ICE, Universidade Federal de Juiz de Fora,
MG, Brazil} 

\date{\today}

\begin{abstract}
In the recent work [1], authors discussed the relationship between the two of Euler angles assuming that it can be used to describe some effects in the theory of a rigid body. I show that this assumption is not properly justified. 
\end{abstract}

\maketitle 




In the recent work \cite{Sug_2020}, authors introduced and discussed a first-order differential equation that relates two of the Eiler angles obeying the Euler-Poisson equations under very special initial conditions, see Eq. (1) in \cite{Sug_2020}.  Since it was obtained from the equations, 
that can be used to study the dynamics of a free asymmetric rigid body, authors assumed that their equation is suitable for describing some effects in the theory of a rigid body, including the Tennis Racket, Dzhanibekov and Monster Flip effects. However, this assumption is not justified, and the relationship of their solution $\psi(\varphi)$ to Eq. (1) in \cite{Sug_2020} with the motions of a rigid body is not clear.  

To confirm this, I present below the analytic solution to the rigid body equations of motion under the assumed by authors initial conditions, and show that properties of this solution are different from those predicted in \cite{Sug_2020}  with help of the function $\psi(\varphi)$. 

The motions of a free rigid body in the center of mass system are described by the Hamiltonian system of Euler-Poisson equations \cite{Eul_1758,Lag_1788,Poi_1842,Hol_1998,Mar_1990,AAD23}
\begin{eqnarray}\label{t1} 
\dot R_{ij}=-\epsilon_{jkp}\Omega_k R_{ip}, \quad 
I\dot{\boldsymbol\Omega}=[I{\boldsymbol\Omega},{\boldsymbol\Omega}]. 
\end{eqnarray}
Here the orthogonal matrix $R_{ij}$ describe the rotational degrees of freedom, $\Omega_i$ is the Hamiltonian counterpart of angular velocity in the body, and the inertia tensor is taken in the diagonal form, $I=diagonal (I_1, I_2, I_3)$.  The integrals of motions $E=\frac12 I_i\Omega_i^2$ 
and ${\bf m}=RI{\boldsymbol\Omega}$, being consequences of these equations, can be added to the system. Then we can omite the Euler equations, since they are consequences of the conservation of the angular momentum ${\bf m}$. Further, using the conservation of momentum in the form $\Omega_k({\bf m})=(I^{-1}R^T{\bf m})_k$ in the remaining equations, we get the following equations for $R_{ij}(t)$, that contain four integration constants $E$, $m_i$:  
\begin{eqnarray}\label{t3}  
\dot R_{ij}=-I_k^{-1}\epsilon_{jkp}(m_1R_{1k}+m_2R_{2k}+m_3R_{3k}) R_{ip},  \quad 
E=\frac12 I_i\Omega_i^2({\bf m}).  
\end{eqnarray}
Given solution $R_{ij}(t)$ to these equations, the dynamics of the point ${\bf x}_N(t)$ of the rigid body is 
${\bf x}_N(t)=R(t){\bf x}_N(0)$. 
This implies that the problem (\ref{t3}) should be solved with the universal initial conditions: $R_{ij}(0)=\delta_{ij}$. When we work with rigid body in terms of Eiler angles, this implies $\theta(t)\rightarrow 0$ as $t\rightarrow 0$. 
Therefore, not all solutions to the equations (\ref{t3}) describe possible motions of a body, but only those passing through the unit element of $SO(3)$ at some instant of time. As a consequence, the initial dates for the angular velocity cannot be taken arbitrary, but are fixed by the values of conserved angular momentum: 
$m_i=I_{ij}\omega_j(0)=I_{ij}\Omega_j(0)$.  
Besides, on this subset of solutions the integration constants are not independent
$2E= I^{-1}_i m_i^2$,
see \cite{AAD23} for the details. 

The equation (1) of the work \cite{Sug_2020} is a consequence of the Euler-Poisson equations, in which the value of conserved angular momentum is chosen as ${\bf m}=(0, 0, m_3)$, see the discussion below Eq. (1), and Eqs. (7) in \cite{Sug_2016}.  
With this angular momentum, the equations of motion (\ref{t3}) reduce to 
\begin{eqnarray}\label{t7.1}
\dot R_{ij}=-\frac{m_3}{I_k}\epsilon_{jkp}R_{3k}R_{ip}. 
\end{eqnarray}
But then the (unique) solution to these equations is obvious, and is as follows:
\begin{eqnarray}\label{t8}
R= \left(
\begin{array}{ccc}
\cos\omega t &  -\sin\omega t & 0  \\
\sin\omega t & \cos\omega t &  0 \\
0 & 0 & 1
\end{array}\right). 
\end{eqnarray}
That is the body fixed frame rotates around the laboratory axis $z$ with constant angular velocity $\omega=m_3/I_3$. The same can be confirmed geometrically, by analyzing the picture of motion according to Poinsot, see Sect. IX in \cite{AAD23} for the details.  

Definitely, this solution does not have the properties of the tennis racket, that were predicted for it in the work \cite{Sug_2020} on the base of analysis  of the function $\psi(\varphi)$, that authors associated with this solution. 

In this regard, I emphasize that for the general choice of conserved angular momentum, the Euler-Poisson equations are much more complicated, see (\ref{t3}).
They can not be reduced to the simple form (\ref{t7.1}) by a suitable rotation of the Laboratory basis.  
Indeed, when writing out the equations (\ref{t1}), it is assumed that at initial instant the laboratory and rigid body axes were chosen in the direction of inertia axes. Due to this, the tensor of inertia is a diagonal matrix, and we deal with rather simple 
expression $\Omega_k({\bf m})=m_3R_{3k}/I_k$, if the momentum is in the direction of $z$\,-axis.   
If we consider a rigid body with an arbitrary angular momentum, and try to rotate the Laboratory system making $z$\,-axis to be collinear with ${\bf m}$, the diagonal matrix $I$ turn into a symmetric matrix $I'$, and we will still be dealing with the complicated angular 
velocity:  $\Omega_k({\bf m})=m_3R_{3i}I'^{-1}_{ik}$. 

To better clarify these issues, let me consider the situation in terms of Euler angles, as was done in \cite{Sug_2020}.
In terms of Euler angles, the solution (\ref{t8}) should have the following structure: $(\varphi(t), \theta(t)=0, \psi(t))$. That is  it lies outside the Euler coordinate system \cite{Arn_1, AAD23}. 
So one cannot expect that the solution could be found by solving our equations with use of the Euler coordinates. 
Let's see what happens, if we nevertheless try to do this.  Writting down equations (\ref{t3}) in the Euler coordinates and 
imposing ${\bf m}=(0, 0, m_3)$, we get \cite{AAD23} 
\begin{eqnarray}\label{t9}
\dot\theta=m_3 I_{(1-2)}\sin\theta\sin\psi\cos\psi, \cr
\dot\varphi=\frac{m_3}{I_1}\sin^2\psi+\frac{m_3}{I_2}\cos^2\psi,  \cr
\dot\psi=-(\frac{m_3}{I_1}\sin^2\psi+\frac{m_3}{I_2}\cos^2\psi)\cos\theta+\frac{m_3}{I_3}\cos\theta. \\
2E=\frac{m^2_3}{I_3}+m^2_3[I_{(1-3)}-I_{(1-2)}\cos^2\psi]\sin^2\theta, \label{t10}
\end{eqnarray}
where $I_{(n-k)}=\frac{1}{I_n}-\frac{1}{I_k}$.  
The Eqs. (\ref{t9}) coincide with Eqs. (7) of the work \cite{Sug_2016}. They imply   
\begin{eqnarray}\label{t11}
\frac{d\psi}{d\varphi}=\pm
\frac{\sqrt{(I_{(1-3)}-I_{(1-2)}\cos^2\psi)(1/I_3-2E/m_3^2+I_{(1-3)}-I_{(1-2)}\cos^2\psi)}}{1/I_1-I_{(1-2)}\cos^2\psi}.
\end{eqnarray}
This is the basic equation (1) studied in \cite{Sug_2020}. If the solution we are looking for describes the motion of a rigid body, it must satisfy the abovementioned relation between the integration constants $E$ and $m_i$. Using it  in Eq. (\ref{t10}), we obtain the following equation: 
$[I_{(1-3)}-I_{(1-2)}\cos^2\psi]\sin^2\theta=0$.
Since we work in vicinity of a point with $\theta\ne 0$, this equation implies that the angle $\psi$ does not change with time:
$\psi=\psi_0$, such that $\cos^2\psi_0=[I_2(I_3-I_1)]/[I_3(I_2-I_1)]$.
By the way, as one might expect, this $\psi_0$ represents a particular solution to (\ref{t11}). 
The last equation of the system (\ref{t9}) is satisfied by this $\psi(t)=\psi_0$, while the remaining two equations read: 
$\dot\theta=\frac12 m_3I_{(1-2)}\sin2\psi_0 \sin\theta$, 
$\dot\varphi=\frac{m_3}{I_3}$, 
and can be immediately integrated 
\begin{eqnarray}\label{t15}
\cos\theta(t)=\frac{1-ce^{2kt}}{1+ce^{2kt}}, \quad 
\varphi=\frac{m_3}{I_3}t+\varphi_0,   \quad 
\psi=\psi_0. 
\end{eqnarray}
Here $c> 0$, and $k\equiv\frac12 m_3I_{(1-2)}\sin2\psi_0\ne 0$. Once again, these solutions to the Euler-Poisson equations, that obeys also the equation (\ref{t11}), have no properties of the tennis racket. 

Moreover, none of the solutions (\ref{t15}) describes the motion of a rigid body. Indeed, we are interested in the solutions with the property $\lim_{t\rightarrow t_0}\theta(t)=0$ for some finite value $t_0$. For any value of the integration constant $c$, there is no such $t_0$. 

In resume,  the solution (\ref{t8}) of Euler-Poisson equations,  with which the authors of the work \cite{Sug_2020}    
associated their  function $\psi(\varphi)$, does not have the properties of this function. 
Therefore, it is not yet clear whether the properties of this function have anything to do with the motions of an asymmetric rigid body.

\vspace{5mm}
\noindent
{\bf Acknowledgments.} The work has been supported by the Brazilian foundation CNPq (Conselho Nacional de
Desenvolvimento Cient\'ifico e Tecnol\'ogico - Brasil). 

\end{document}